\documentclass[12pt]{article}
\usepackage{makeidx}
\usepackage{amssymb}
\usepackage{amsfonts}
\usepackage{amsmath}
\usepackage{graphicx}
\usepackage{setspace}
\usepackage{authblk}
\usepackage[left=1.6cm,top=2cm,right=1.6cm]{geometry}
\usepackage{units}

\begin{document}

\title{\textbf{Electric charge is a magnetic dipole when placed in a background magnetic field}}

\author[1,2]{T. C. Adorno\thanks{tadorno@usp.br, tadorno@ufl.edu}}
\author[1]{D. M. Gitman\thanks{gitman@dfn.if.usp.br}}
\author[3]{A. E. Shabad\thanks{shabad@lpi.ru}}
\affil[1]{\textit{Instituto de F\'{\i}sica, Universidade de S\~{a}o Paulo, Caixa Postal 66318, CEP 05508-090, S\~{a}o Paulo, S.P., Brazil;}}
\affil[2]{\textit{Department of Physics, University of Florida, 2001 Museum Road, Gainesville, FL 32611-8440, USA;}}
\affil[3]{\textit{P. N. Lebedev Physics Institute, 117924 Moscow, Russia.}}

\maketitle

\begin{abstract}
It is demonstrated, owing to the nonlinearity of QED, that a static charge
placed in a strong magnetic field\ $B$\ is a magnetic dipole (besides
remaining an electric monopole, as well). Its magnetic moment grows linearly
with $B$ as long as the latter remains smaller than the characteristic value
of $1.2\cdot 10^{13}\unit{G}$ but tends to a constant as $B$ exceeds that
value. The force acting on a densely charged object by the dipole magnetic
field of a neutron star is estimated.
\end{abstract}

\newpage

\section{Introduction}

In the recent papers \cite{ShaGit2012,CGS}, we started a study of
self-interaction of static electric and magnetic fields of moderate
strength -- in the vacuum \cite{CGS} or taken against a very strong
background formed by a constant and homogeneous magnetic field \cite%
{ShaGit2012}. By "moderate" we mean the fields that are strong enough to
make the nonlinearity of QED actual for their
self-interaction, but still assumed smaller than the characteristic value $B_{\text{Sch}}=m^{2}/e,$ where $m$ and $e$ are electron mass and charge, to
enable us to exploit the expansion of the nonlinear Maxwell equations in
powers of the fields and confine ourselves to the lowest terms in this
expansion. On the contrary, the background field is not limited either from
below or from above. It was observed that effects of selfinteraction
manifest themselves already within the simplest approximation of local
action, valid in the infrared region of momenta, i.e. for fields that do not
change any essentially at the time or space interval of $m^{-1}.$ Among
these effects are nonlinear corrections to the Coulomb field in the vacuum 
\cite{CGS}, making the field energy of a point charge finite \cite{Caio2},
electromagnetic nonlinear renormalization of electric and magnetic dipole
moments of mesons and baryons \cite{CGS}, necessary after the latter are
calculated following the theory of strong interaction. In the present paper,
we continue the work \cite{ShaGit2012}, where it was pointed out that the
quadratic response of the strong background magnetic field to an applied
moderate electric field is purely magnetic, and the nonlinear current
induced by an applied electric field \ was calculated based on the
third-rank polarization tensor (a three-photon vertex beyond the mass shell)
in the infrared (local) limit. Also, a general expression was given for the
magnetic field nonlinearly induced by an electrostatic field. Below we 
demonstrate that this field is that of a magnetic dipole. The magnetic
dipole moment carried by a static spherical charge of finite extension is
calculated. We estimate the force acting on this magnetic moment by the
inhomogeneous magnetic field of a neutron star and a magnetar. If we admit
that in the course of the phase transition of the neutron star to a quark
star \cite{Negreiros,quark stars} highly charged domains of strange
matter \cite{strangelets} may exist, then the observed effect may form a
mechanism of charge outflow from the star.

\section{Induced current}

The nonlinear response of the magnetized vacuum to applied electrostatic
field within the nonlinear Maxwell equations truncated at the third power of
the field is purely magnetic \cite{ShaGit2012}, with the vector potential
components $a_{\mathrm{nl}}^{\mu }$ being given as (we keep to the
rationalized Heaviside-Lorentz units throughout) 
\begin{equation}
a_{\mathrm{nl}}^{0}=0\,,\ \ a_{\text{\textrm{nl}}}^{i}(k)=\sum_{c=1,3}\frac{%
2\pi \delta (k_{0})}{(k^{2}-\varkappa _{c}(k))}\frac{\flat _{i}^{(c)}}{%
(\flat ^{(c)})^{2}}(\widetilde{j}_{j}^{\mathrm{nl}}\left( \mathbf{k}\right)
\flat _{j}^{(c)})\,,  \label{1}
\end{equation}%
where $\widetilde{j}_{j}^{\mathrm{nl}}\left( \mathbf{k}\right) $ is the
Fourier-transformed nonlinearly induced current%
\begin{equation}
j_{\rho }^{\mathrm{nl}}\left( x\right) =-\frac{1}{2}\int d^{4}x^{\prime
}d^{4}x^{\prime \prime }\Pi _{\nu \sigma \rho }\left( x^{\prime },x^{\prime
\prime },x\right) a^{\nu }\left( x^{\prime }\right) a^{\sigma }\left(
x^{\prime \prime }\right) \,,\,  \label{nonlinear current1}
\end{equation}%
and $\Pi _{\nu \sigma \rho }\left( x^{\prime },x^{\prime \prime },x\right) $
is the third-rank polarization tensor in the coordinate representation
defined as the third variational derivative of the effective action $%
\mathcal{\Gamma }$ with respect to the field potential taken in three
space-time points $x^{\prime },x^{\prime \prime },x.$ In QED it includes all
diagrams with three off-shell outer photons in the external magnetic field.
In (\ref{1}) $\flat _{i}^{(c)}$ are the three eigenvectors of the
second-rank polarization tensor in the background magnetic field,
responsible for the linear response in that field, and $\varkappa _{c}(k)$
are the corresponding eigenvalues. Out of the three components of the photon
propagator that build a field from a (nonlinear) current, only two
contribute into the sum in (\ref{1}). This means that in the static limit
only two eigenmodes are responsible for carrying magnetic field.

If the effective action $\ \Gamma =\int \mathcal{L}\left( x\right) d^{4}x$
is assumed to be a local functional of the scalar $\left( \mathfrak{F}%
\right) $ and pseudoscalar $\left( \mathfrak{G}\right) $ field invariants,
like what the Euler-Heisenberg in QED, or Born-Infeld actions are, the
current (\ref{nonlinear current1}) has the form \cite{ShaGit2012}:%
\begin{equation}
j_{\mathrm{nl}}^{0}\left( \mathbf{x}\right) =0\,,\ \ \mathbf{j}_{\mathrm{nl}%
}\left( \mathbf{x}\right) =\mathbf{\nabla \times }\text{ }\mathfrak{h}\left( 
\mathbf{x}\right) ,  \label{nonlinear current2}
\end{equation}

\begin{equation}
\mathfrak{h}_{i}\left( \mathbf{x}\right) =\frac{B_{i}}{2}\mathcal{L}_{%
\mathfrak{FF}}\mathbf{E}^{2}-\frac{B_{i}}{2}\mathcal{L}_{\mathfrak{FGG}%
}\left( \mathbf{B}\cdot \mathbf{E}\right) ^{2}-\mathcal{L}_{\mathfrak{GG}%
}\left( \mathbf{B}\cdot \mathbf{E}\right) E_{i}\,.  \label{auxiliary field}
\end{equation}%
Here $\mathbf{E}$ is the static applied field contained in the
vector-potential in the right-hand side of ( \ref{nonlinear current1}), and $%
\mathbf{B}$ is the external constant and the homogeneous magnetic field
contained in $\Pi _{\nu \sigma \rho }\left( x^{\prime },x^{\prime \prime
},x\right) .$ The scalar coefficients in the auxiliary field (\ref{auxiliary
field}) are the derivatives of the effective Lagrangian $\mathcal{L}\left(
x\right) $ with respect to the field invariants taken at $\mathfrak{G=}$ $0,$
$2\mathfrak{F=}$ $B^{2}$ = \emph{const.} They depend only on $B=|\mathbf{B}%
|. $

\section{Induced magnetic field and magnetic moment}

The magnetic field strength $\boldsymbol{h}\mathbf{(x)}$ generated by the
current (\ref{nonlinear current2})\ according to the Maxwell equation $%
\boldsymbol{\nabla }\mathbf{\times }\boldsymbol{h}\mathbf{(x)=}$ $%
\boldsymbol{j}^{\text{nl}}(\mathbf{x})$ is 
\begin{equation}
h_{i}\mathbf{(x)=}\mathfrak{h}_{i}(\mathbf{x)+}\nabla _{i}\Omega ,
\label{magnstrength-1}
\end{equation}%
because $\boldsymbol{\nabla }\mathbf{\times }\boldsymbol{\nabla }\Omega
\equiv 0.$ To find the scalar function $\Omega $ \ we should exploit the
other Maxwell equation $\left( \mathbf{\nabla \cdot b}\right) =0,$ where the
magnetic induction $\mathbf{b}$\ is related to the magnetic field as $h_{i}%
\mathbf{(x)=}\mu _{ij}^{-1}b_{j}$ through the (inverse) magnetic
permeability tensor of the magnetized vacuum, in the present local
approximation given as \cite{villalbachaves} $\mu _{ij}^{-1}=\left( 1-%
\mathfrak{L}_{\mathfrak{F}}\right) \delta _{ij}-\mathfrak{L}_{\mathfrak{F}%
\mathfrak{F}}B_{i}B_{j}.$ Its eigenvalues are $\mu _{\perp }^{-1}=$ $1-%
\mathfrak{L}_{\mathfrak{F}},$ $\mu _{\parallel }^{-1}=$ $1-\mathfrak{L}_{%
\mathfrak{F}}-\mathfrak{L}_{\mathfrak{F}\mathfrak{F}}B^{2}.$ Then 
\begin{equation*}
\Omega =-\frac{\nabla _{i}\mu _{ij}}{\nabla _{m}\mu _{mn}\nabla _{n}}%
\mathfrak{h}_{j}(\mathbf{x)};
\end{equation*}%
hence (the tilde designates again the Fourier transform),%
\begin{eqnarray}
h_{i}\mathbf{(x)} &\mathbf{=}&\left( \delta _{ij}-\nabla _{i}\frac{\nabla
_{j}+\mathfrak{L}_{\mathfrak{F}\mathfrak{F}}\mu _{\parallel }\left( \mathbf{%
B\cdot \nabla }\right) B_{j}}{\boldsymbol{\nabla }^{2}+\mathfrak{L}_{%
\mathfrak{F}\mathfrak{F}}\mu _{\parallel }\left( \mathbf{B\cdot \nabla }%
\right) ^{2}}\right) \mathfrak{h}_{j}(\mathbf{x)}  \label{magnstrength1} \\
&\mathbf{=}&\text{ }\mathfrak{h}_{i}(\mathbf{x)+}\nabla _{i}\frac{\nabla
_{j}+\mathfrak{L}_{\mathfrak{F}\mathfrak{F}}\mu _{\parallel }\mathbf{(B\cdot
\nabla )}B_{j}}{(2\pi )^{3}}\int \frac{\widetilde{\mathfrak{h}}_{j}(%
\boldsymbol{k})e^{i\mathbf{k}\cdot \mathbf{x}}}{\boldsymbol{k}^{2}+\mathfrak{%
L}_{\mathfrak{F}\mathfrak{F}}\mu _{\parallel }\left( \mathbf{B\cdot k}%
\right) ^{2}}d^{3}k,
\end{eqnarray}%
and 
\begin{eqnarray}
h_{i}\left( \mathbf{x}\right) &=&\mathfrak{h}_{i}\left( \mathbf{x}\right)
+\nabla _{i}\left[ \nabla _{k}+\left( \frac{\mu _{\parallel }}{\mu _{\perp }}%
-1\right) \frac{\mathbf{(B\cdot \nabla )}B_{k}}{B^{2}}\right] \mathfrak{T}%
_{k}\left( \mathbf{x}\right) ,\ \   \label{magneticstrength2} \\
\mathfrak{T}_{k}\left( \mathbf{x}\right) &=&\frac{1}{4\pi }\int d^{3}y\frac{%
\mu _{\perp }^{1/2}\mathfrak{h}_{k}\left( \mathbf{y}\right) }{\left( \mu
_{\parallel }\left( \mathbf{x}_{\perp }\mathbf{-y}_{\perp }\right) ^{2}+\mu
_{\perp }\left( \mathbf{x}_{_{\parallel }}\mathbf{-y}_{\parallel }\right)
^{2}\right) ^{1/2}}\,.  \label{teka}
\end{eqnarray}%
The indices $\parallel $ and $\perp $ mark the radius-vector components,
parallel and orthogonal to the background magnetic field $\mathbf{B.}$\ \ We are interested in asymptotic behavior of (\ref{magneticstrength2}) in
the far-off region $x_{i}\rightarrow \infty .$ The contribution of (\ref%
{teka}) is%
\begin{equation*}
\mathfrak{T}_{k}\left( \mathbf{x}\right) \sim \frac{1}{4\pi }\frac{\mu
_{\perp }^{1/2}\int d^{3}y\text{ }\mathfrak{h}_{k}\left( \mathbf{y}\right) }{%
\left( \mu _{\parallel }\mathbf{x}_{\perp }^{2}+\mu _{\perp }\mathbf{x}%
_{\parallel }^{2}\right) ^{1/2}}
\end{equation*}%
provided that the integral here converges. This is guaranteed assuming that
the applied electric field $E_{i}\,$\ in (\ref{auxiliary field}) decreases
fast enough at $y_{i}\rightarrow \infty $. If its source is concentrated in
a finite space domain, $E$ decreases as fast as square of the distance from
this domain, and the auxiliary field (\ref{auxiliary field}) as its fourth
power. This decrease is faster than that of the second term in (\ref%
{magneticstrength2}). Therefore only the latter should be kept in the remote
region%
\begin{eqnarray}
h_{i}\left( \mathbf{x}\right) &\sim &\frac{\mu _{\perp }^{1/2}}{4\pi }\int
d^{3}y\text{ }\mathfrak{h}_{k}\left( \mathbf{y}\right) \nabla _{i}\left(
\nabla _{k}+\left( \frac{\mu _{\parallel }}{\mu _{\perp }}-1\right) \frac{%
\mathbf{(B\cdot \nabla )}B_{k}}{B^{2}}\right) \frac{1}{\left( \mu
_{_{\parallel }}\mathbf{x}_{\perp }^{2}+\mu _{\perp }\mathbf{x}_{\parallel
}^{2}\right) ^{1/2}}=  \notag \\
&=&\frac{\mu _{\perp }^{1/2}\mu _{_{\parallel }}}{4\pi }\int d^{3}y\mathfrak{%
h}_{k}\left( \mathbf{y}\right) \frac{1}{\check{r}^{3}}\left( -\delta _{ki}+%
\frac{3x_{k}x_{i}\mu _{_{\parallel }}}{\check{r}^{2}}+\frac{3x_{k}B_{i}(%
\mathbf{B\cdot x)(}\mu _{\perp }-\mu _{_{\parallel }})}{B^{2}\check{r}^{2}}%
\right) ,  \label{dipole}
\end{eqnarray}%
where we have denoted the \textit{effective} distance as $\check{r}=\left( \mu _{_{\parallel }}%
\mathbf{x}_{\perp }^{2}+\mu _{\perp }\mathbf{x}_{\parallel }^{2}\right)
^{1/2}.$ This magnetic field satisfies both sourceless Maxwell equations 
$\boldsymbol{\nabla }\mathbf{\times }\boldsymbol{h}\mathbf{(x)=}0,$ $\left( 
\mathbf{\nabla \cdot b(x)}\right) =0$ in the whole space domain (not only in
the remote region), except the point $r^{\prime }=0.$ It is the field of a
pointlike magnetic dipole in anisotropic medium, with the magnetic moment
being $\mathbf{M}=\frac{\mu _{\perp }^{1/2}\mu _{_{\parallel }}}{4\pi }\int
d^{3}y$ $\mathfrak{h}\left( \mathbf{y}\right) $. If we also assume that the
electric field is no less than cylindrically symmetric with the symmetry
axis coinciding with the direction of $\mathbf{B,}$ the magnetic moment
becomes parallel with $\mathbf{B.}$ In the limit of isotropic medium $\mu
_{\perp }=\mu _{_{\parallel }}=\mu $ \ the last term in (\ref{dipole})
disappears, and $\check{r}$ becomes $\mu ^{1/2}\left\vert \mathbf{x}\right\vert$; hence, $\mu $ cancels out from Eq. (\ref{dipole}), and the latter
acquires the standard form of a magnetic dipole in the vacuum

\begin{equation}
\mathbf{h}\left( \mathbf{x}\right) =\frac{1}{r^{3}}\left( -\mathbf{M}+\frac{3%
\mathbf{x}(\mathbf{x\cdot M})}{r^{2}}\right) ,\text{ \ \ }\mathbf{M}=\frac{1%
}{4\pi }\int d^{3}y\mathfrak{h}\left( \mathbf{y}\right) ,  \label{dipole2}
\end{equation}%
where now $r$ is just $\left\vert \mathbf{x}\right\vert .$ Earlier \cite%
{AdoGitSha2013} we reported this result under the neglect of the linear
response of the vacuum by setting $\mu _{\perp }=\mu _{_{\parallel }}=\mu =1$%
. Now we see that it is sufficient to admit only the absence of anisotropy
of this response.\ In the magnetized vacuum in QED, the isotropization occurs 
\cite{ShaGit2012} for very large background field $B\gg 2.7\cdot B_{\text{Sch%
}},$ where $\mathfrak{L}_{\mathfrak{F}}\gg 2\mathfrak{L}_{\mathfrak{F}%
\mathfrak{F}}B^{2}.$ However, if QED is treated perturbatively, the whole
linear response of the background magnetic field introduces only a
higher order correction into (\ref{dipole}): for small $B$ the coefficients\ 
$\mathcal{L}_{\mathfrak{FF}},$ $\mathcal{L}_{\mathfrak{FGG}}\mathbf{,}$ $%
\mathcal{L}_{\mathfrak{GG}}$ are proportional to the fine-structure constant 
$\alpha =e^{2}/4\pi $ squared [see Eq. (\ref{smallB1}) below], as well as $%
\mathcal{L}_{\mathfrak{F}},$ whereas for large $B$ they all are proportional
to $\alpha $ [ consult Eqs. (\ref{largeBregime}) below and the relation $%
\mathcal{L}_{\mathfrak{F}}=\frac{\alpha }{3\pi }\ln \frac{eB}{m^{2}},$ valid
in the large-field regime].

\section{A basic example}

For demonstrative purposes, we accept this approximation in considering
the simplest example of the magnetic response to a spherically symmetric
electric field given by the potential 
\begin{eqnarray}
&&a_{0}\left( r\right) =a_{0}^{\mathrm{in}}\left( r\right) \theta \left(
R-r\right) +a_{0}^{\mathrm{out}}\left( r\right) \theta \left( r-R\right)
\,,\,\,r=\left\vert \mathbf{x}\right\vert  \notag \\
&&a_{0}^{\mathrm{in}}\left( r\right) =-\frac{Ze}{8\pi R^{3}}r^{2}+\frac{3}{%
8\pi }\frac{Ze}{R}\,,\ \ a_{0}^{\mathrm{out}}\left( r\right) =\frac{Ze}{4\pi
r}\,,  \label{1.1}
\end{eqnarray}%
where $\theta \left( z\right) $ is the step function

\begin{equation*}
\theta \left( z\right) =\left\{ 
\begin{array}{c}
1\,,\ \ z>0\,, \\ 
0\,,\ \ z<0\,.%
\end{array}%
\right. \,.
\end{equation*}%
Once the applied electric field is spherically symmetric, $\mathbf{E}=%
\mathbf{x}\mathcal{E}\left( r\right) ,$ the current $\mathbf{j}_{\mathrm{nl}%
}\left( \mathbf{x}\right) $ (\ref{nonlinear current2}), (\ref{auxiliary
field}) is proportional to the vector product $\mathbf{B\times x,}$ hence it
is circular in the plane orthogonal to the background magnetic field.
Simultaneously, the contribution of the first mode $c=1$ disappears from (%
\ref{1}). \ \ \ \ \ \ \ \ \ \ \ \ \ \ \ \ \ \ \ \ \ \ \ \ \ \ \ \ \ \ \ \ \
\ \ \ \ \ \ \ \ \ \ \ \ \ \ \ \ \ \ \ \ \ \ \ \ \ \ \ \ \ \ \ \ \ \ \ \ \ \
\ \ \ \ \ \ \ \ \ \ \ \ \ \ \ \ \ \ \ \ \ \ \ \ \ \ \ \ \ \ \ \ \ \ \ \ \ \
\ \ \ \ \ \ \ \ \ 

\begin{figure}[th!]
\begin{center}
\includegraphics[scale=0.5]{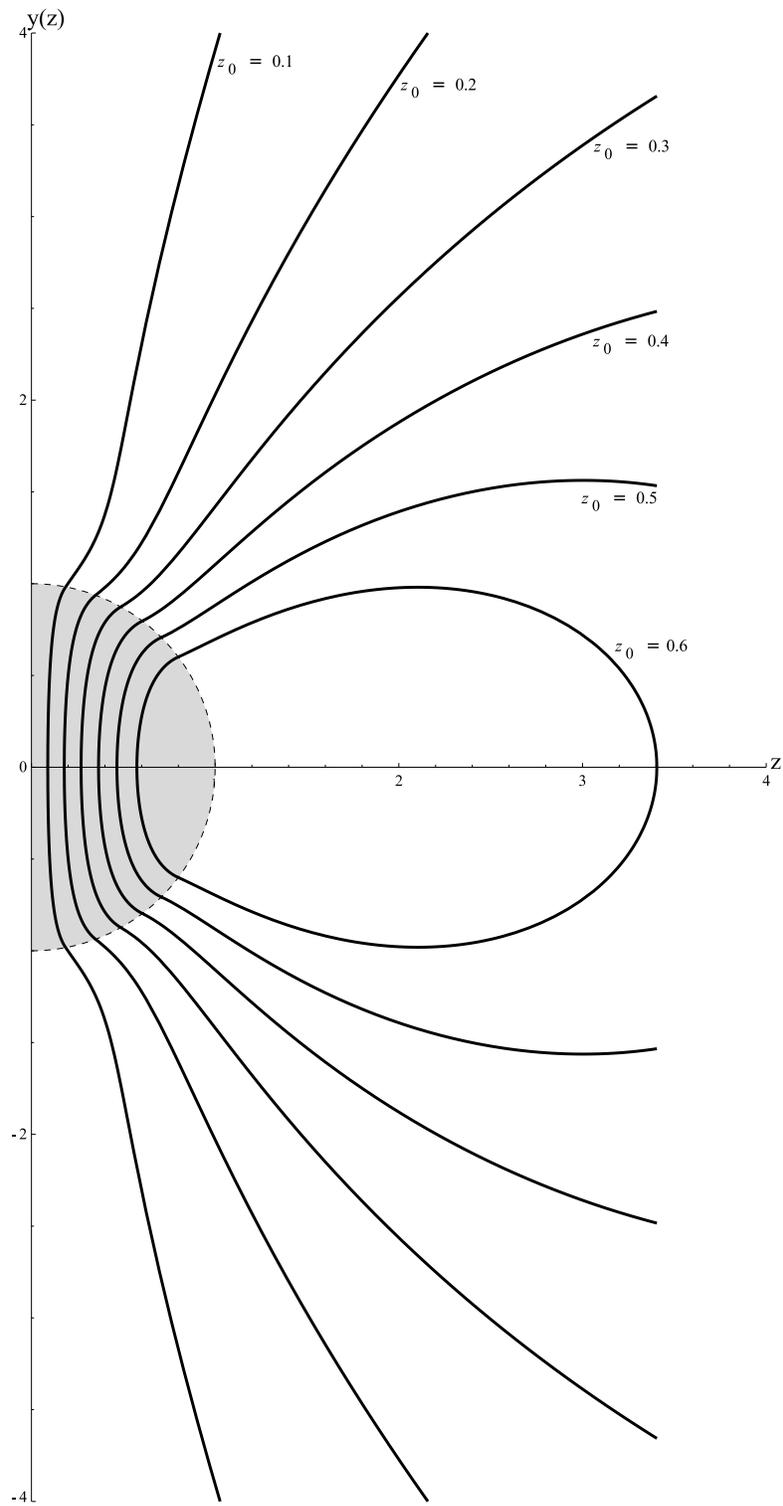}
\end{center}
\caption{Magnetic dipole lines of a static charge in an external magnetic field exampled with $B=\infty $. Shaded is the area of the charge.}
\label{Fig1}
\end{figure}

If not for the linear electric polarization, the potential (\ref{1.1}) would
be the field of extended spherically symmetric charge \bigskip distributed
with the constant density $\rho \left( r\right) $ inside a sphere $r\leq R$
with the radius $R$:%
\begin{equation}
\rho \left( r\right) =\left( \frac{3}{4\pi }\frac{Ze}{R^{3}}\right) \theta
\left( R-r\right) \,.  \label{rho}
\end{equation}%
\ With the account of the linear vacuum polarization, the potential
distribution (\ref{1.1}) cannot be supported by any spherically symmetric
charge, strictly localized in a finite space domain. The genuine source of
the field (\ref{1.1}) is $\rho _{\text{lin}}\left( \mathbf{x}\right) =\left( 
\mathbf{\nabla \cdot d}\right) $, where the electric induction is defined as 
$d_{i}\mathbf{=-}\varepsilon _{ij}\nabla _{j}a_{0}$ in terms of the
potential (\ref{1.1}) and of the dielectric tensor $\varepsilon _{ij}$ of
the vacuum, which in the constant magnetic background has, in the present
local approximation, the form \cite{villalbachaves} $\varepsilon
_{ij}=\left( 1-\mathfrak{L}_{\mathfrak{F}}\right) \delta _{ij}+\mathfrak{L}_{%
\mathfrak{G}\mathfrak{G}}B_{i}B_{j}.$ We find 
\begin{equation*}
\rho _{\text{lin}}\left( \mathbf{x}\right) =\rho \left( r\right) (1-\mathcal{%
L}_{\mathfrak{F}})+2\mathfrak{F}\mathcal{L}_{\mathfrak{GG}}\left( 1+\frac{(%
\mathbf{B\cdot x})^{2}}{B^{2}}\frac{d}{rdr}\right) \frac{d}{rdr}a_{0}\left(
r\right) .
\end{equation*}%
This charge density is cylindrically symmetric and extends beyond the
sphere, $r>R,$ decreasing as 1/$r^{3},$ or 1/$x_{3}^{3}$ far from it,
depending on the direction. However, the same argument as above allows us to
neglect the linear polarization in the electric sector by setting $%
\varepsilon _{ij}=\delta _{ij}$, as well as we did in the magnetic. Then $%
\rho _{\text{lin}}\left( \mathbf{x}\right) =\rho \left( r\right) ,$ and
therefore, we refer to the magnetic field nonlinearly produced by the
electric field (\ref{1.1}) as the nonlinear response to the field of a
homogeneously charged sphere (\ref{rho}). It is calculated following Eqs. (%
\ref{magneticstrength2}) and (\ref{teka}) with $\mathbf{E=-\nabla }a_{0}$ used
in (\ref{auxiliary field}) and with $\mu _{_{\parallel }}=\mu _{\perp }=1$
to be (the details of this calculation can be found in \cite{AdoGitSha2013}$)
$ 
\begin{eqnarray}
h_{i}\left( \mathbf{x}\right)  &=&h_{i}^{\mathrm{in}}\left( \mathbf{x}%
\right) \theta \left( R-r\right) +h_{i}^{\mathrm{out}}\left( \mathbf{x}%
\right) \,\theta \left( r-R\right) ,  \label{total} \\
h_{i}^{\mathrm{in}}\left( \mathbf{x}\right)  &=&-\left( \frac{Ze}{4\pi R^{2}}%
\right) ^{2}\left\{ \left[ \frac{1}{2}\left( 1-\frac{4r^{2}}{5R^{2}}\right) 
\mathcal{L}_{\mathfrak{GG}}+\frac{r^{2}}{10R^{2}}\mathcal{L}_{\mathfrak{FF}%
}\right. \right.   \notag \\
&+&\left. \mathcal{L}_{\mathfrak{FGG}}\left( \frac{1}{10}\left( 1-\frac{%
4r^{2}}{7R^{2}}\right) B^{2}+\frac{2r^{2}}{7R^{2}}\left( \frac{\mathbf{%
B\cdot x}}{r}\right) ^{2}\right) \right] B_{i}  \notag \\
&+&\left. \frac{1}{5}\left( \frac{1}{2}\mathcal{L}_{\mathfrak{FF}}+\frac{1}{2%
}\mathcal{L}_{\mathfrak{GG}}-\frac{2}{7}B^{2}\mathcal{L}_{\mathfrak{FGG}%
}\right) \frac{\left( \mathbf{B}\cdot \mathbf{x}\right) x_{i}}{r^{2}}%
\right\} ,  \notag \\
h_{i}^{\mathrm{out}}\left( \mathbf{x}\right)  &=&\left( \frac{Ze}{4\pi r^{2}}%
\right) ^{2}\left\{ \left[ \frac{1}{2}\left( 1-\frac{6r}{5R}\right) \mathcal{%
L}_{\mathfrak{FF}}-\frac{1}{2}\left( 1-\frac{4r}{5R}\right) \mathcal{L}_{%
\mathfrak{GG}}\right. \right.   \notag \\
&-&\left. \left( \frac{1}{2}\left( 1-\frac{2r}{5R}-\frac{18R}{35r}\right)
B^{2}-\left( 1-\frac{9R}{7r}\right) \left( \frac{\mathbf{B\cdot x}}{r}%
\right) ^{2}\right) \mathcal{L}_{\mathfrak{FGG}}\right] B_{i}  \notag \\
&-&2\left( \frac{Ze}{4\pi r^{2}}\right) ^{2}\left\{ \left( 1-\frac{9r}{10R}%
\right) \mathcal{L}_{\mathfrak{FF}}-\frac{1}{2}\left( 1-\frac{6r}{5R}\right) 
\mathcal{L}_{\mathfrak{GG}}\right.   \notag \\
&+&\left. \left[ \left( -1+\frac{3r}{10R}+\frac{9R}{14r}\right) B^{2}+\frac{3%
}{2}\left( 1-\frac{R}{r}\right) \left( \frac{\mathbf{B\cdot x}}{r}\right)
^{2}\right] \mathcal{L}_{\mathfrak{FGG}}\right\} \frac{\left( \mathbf{B}%
\cdot \mathbf{x}\right) x_{i}}{r^{2}}\,.  \notag
\end{eqnarray}%
The shape of a family of magnetic lines of force is depicted in Fig. \ref{Fig1}
drawn, for definiteness, for asymptotically large values of the background
field $B\gg m^{2}/e,$ in which domain it holds, using the Euler-Heisenberg
effective Lagrangian (see e.g. \cite{Shabus2011}), that

\begin{equation}
\mathcal{L}_{\mathfrak{FF}}=\frac{\alpha }{3\pi }\frac{1}{B^{2}}\,,\text{\ \ 
}\mathcal{L}_{\mathfrak{GG}}=\frac{\alpha }{3\pi }\left( \frac{e}{m^{2}}%
\right) \frac{1}{B}\,,\text{\ \ }B^{2}\mathcal{L}_{\mathfrak{FGG}}=B^{2}%
\frac{d\mathcal{L}_{\mathfrak{GG}}}{d\mathfrak{F}}=-\mathcal{L}_{\mathfrak{GG%
}}\,.  \label{largeBregime}
\end{equation}%
Inside the charged sphere $R$ the curves follow the formula%
\begin{equation}
y\left( z\right) =\sqrt{\frac{14}{11}-\frac{6}{11}z^{2}-\left( \frac{z_{0}}{z%
}\right) ^{2}}\,,
\end{equation}%
being labeled by positive values of the integration constant $z_{0}$ in the
interval $0<$ $z_{0}<\frac{7}{\sqrt{66}},$ whereas outside the sphere they
have resulted from the computer solution of the corresponding first-order
differential equation $\frac{\text{d}z}{\text{d}y}=\frac{h_{z}}{h_{y}}$ (we
have directed axis $z$ along the background magnetic field, and axis $y$
along any direction in the orthogonal plane). It is seen that in the
long-range domain the pattern of the lines of force is that of a magnetic
dipole. Indeed, the long-range contribution of (\ref{total}) does behave
like a magnetic field of a solenoid (\ref{dipole2}) with the equivalent
magnetic moment $M$ given by%
\begin{equation}
M_{i}=\left( \frac{Ze}{4\pi }\right) ^{2}\frac{1}{5R}\left( 3\mathcal{L}_{%
\mathfrak{FF}}-2\mathcal{L}_{\mathfrak{GG}}-B^{2}\mathcal{L}_{\mathfrak{FGG}%
}\right) B_{i}\,,  \label{magnmoment}
\end{equation}%
which agrees with (\ref{dipole2}), once the auxiliary field (\ref{auxiliary
field}) is taken on the Coulomb field (\ref{1.1}).

\section{Estimates}

To estimate the effect of nonlinear magnetization numerically note that,
within the Euler-Heisenberg Lagrangian, for the background field $B\ll \frac{%
m^{2}}{e}$, much smaller than Schwinger's optional value $B_{\text{Sch}}=%
\frac{m^{2}}{e}=1.2\cdot 10^{13}\unit{G},$ it holds%
\begin{equation}
\mathcal{L}_{\mathfrak{FF}}=\frac{16\alpha ^{2}}{45m^{4}}\text{ \ \ , \ \ }%
\mathcal{L}_{\mathfrak{GG}}=\frac{28\alpha ^{2}}{45m^{4}},  \label{smallB1}
\end{equation}%
while the contribution of the term $B^{2}\mathcal{L}_{\mathfrak{FGG}}$ in (%
\ref{magnmoment}) and in (\ref{total}) is negligible in this order as being
proportional to an extra power of $B^{2}.$ Then in this field domain, after
the substitution $E=-\frac{Ze}{4\pi R^{2}},$ we have from (\ref{total}) for
the component orthogonal to the background field, taken approximately at $%
r=R,$%
\begin{equation}
h_{\perp }=\frac{-11\alpha }{225}\left( \frac{eE}{m^{2}}\right) ^{2}B\frac{%
\sin 2\theta }{2}=3.5\cdot 10^{-4}\left( \frac{E}{E_{\text{Sch}}}\right)
^{2}B\frac{\sin 2\theta }{2},  \label{smallB}
\end{equation}%
where $\theta $ is the angle between $\mathbf{B}$ and the radius-vector $%
\mathbf{x}$, directed toward the points where the induced magnetic field is
orthogonal to $\mathbf{B}$ (see Fig. \ref{Fig1}), and $E_{\text{Sch}}=\frac{m^{2}}{e}%
=1.3\cdot 10^{16}\unit{V}/\unit{cm}$. Our results are not expected to be
applicable to charged microscopic objects, since their electric field close
to themselves is too large to be treatable using only the second-power
nonlinearity (\ref{nonlinear current1}), and varies too fast at the distance
of the electron Compton length $m^{-1}$ to be treatable within the
approximation of the local action. For this reason we consider a macroscopic
device at extremum laboratory conditions. Let the potential of accelerator
scale, $(1\div 25)$ $\unit{MV}$, be applied to a point ball with the
curvature radius $R=\left( 1\div 0.1\right) \cdot 10^{-5}\unit{cm},$ like
the one in an ion projector, to produce electric field $E=\left( 1\div
250\right) \cdot 10^{11}\unit{V}/\unit{cm}=(0.77\div 200)\cdot 10^{-5}E_{%
\text{Sch}}$. Then, according to (\ref{smallB}), $h_{\perp }\simeq (2\cdot
10^{-5}\div 1.4)\cdot 10^{-9}B.$ This quantity may reach maximum values for
the background field that may achieve $B=10^{6}\unit{G}$ in a laboratory. It
is hard to say whether they can be registered against the strong background
field, although directed orthoganally to $h_{\perp }$ .As Eq. (\ref{smallB})
remains valid up to the values of pulsar scale $B\simeq 0.1B_{\text{Sch}%
}=1.2\cdot 10^{12}\unit{G},$ the field of the above device would make $%
h_{\perp }\simeq (2.2\cdot 10^{-5}\div 1.7)\cdot 10^{3}\unit{G}$ if \
placed into such pulsar. The magnetic moment (\ref{magnmoment}) in the
small-field domain follows from (\ref{smallB1}) to be%
\begin{equation}
M\simeq -\frac{Z^{2}\alpha ^{2}}{(Rm)}\frac{1}{225\pi ^{2}}\mu _{e}\frac{eB}{%
m^{2}}\simeq -\left( \frac{E}{E_{\text{Sch}}}\right) ^{2}\frac{2\alpha }{%
225\pi }BR^{3},  \label{magnmomentsmall}
\end{equation}%
where $\mu _{e}=\frac{e}{2m}=9.27\cdot 10^{-21}\unit{G}\cdot \unit{cm}^{3}$
is the Bohr magneton. The force, acting to it by the inhomogeneous (dipole)
magnetic field $B$ near a pulsar surface is $F\simeq M\frac{\text{d}B}{\text{%
d}r}\simeq M\frac{\text{d}}{\text{d}r}\frac{M_{\text{pl}}}{r^{3}}\simeq -3M%
\frac{M_{\text{pl}}}{r^{4}}\simeq -4\pi M\frac{B}{r},$ where $M_{\text{pl}}=B%
\frac{4}{3}\pi r^{3}$ is the magnetic moment of the neutron star, and $r$ is
its radius. If we imagine a macroscopic ball of the radius $R=4\cdot
10^{9}m^{-1}\simeq 1\unit{cm}$ carrying the charge $Ze=$ $1\unit{C}$, i.e., $%
Z=0.6\cdot 10^{19}$ and place it in the magnetic field of a pulsar, taking $%
r=10^{6}\unit{cm}$ for its radius and $B\simeq 0.1B_{\text{Sch}}$ for its
field at the surface, we see that its magnetic moment is $0.2\unit{G}%
\cdot \unit{cm}^{3}$ and it is subject to a force of the order of $\simeq 3$%
\emph{\ }kG. Note that since $r\gg R,$ the magnetic moment has enough space
to be formed before it would sense the inhomogenuity of the background field
and undergo its forcing influence.

In the opposite asymptotic regime $B\gg \frac{m^{2}}{e},$ characteristic of
some magnetars \cite{magnetars}, it follows from (\ref{largeBregime}) that
the second and the third terms in (\ref{magnmoment}) dominate to provide the
saturation of the magnetic moment in the limit $B=\infty $ at the level of%
\begin{equation}
M=-\left( \frac{Ze}{4\pi }\right) ^{2}\frac{\alpha }{15R\pi }\frac{e}{m^{2}}%
=-\frac{Z^{2}\alpha ^{2}}{(Rm)}\frac{1}{30\pi ^{2}}\mu _{e}=-\left( \frac{E}{%
E_{\text{Sch}}}\right) ^{2}\frac{\alpha }{15\pi }B_{\text{Sch}}R^{3},
\label{infiniteB}
\end{equation}%
where $\mu _{e}=\frac{e}{2m}=9.27\cdot 10^{-21}\unit{G}\cdot \unit{cm}^{3}$
is the Bohr magneton$.$ Once $\frac{\alpha }{15\pi }=1.5\cdot 10^{-4},$ the
upper bound on (\ref{infiniteB}) admitted by the quadratic approximation (%
\ref{nonlinear current1}), which requires that $E<E_{\text{Sch}},$ is $\left\vert M\right\vert$ $<1.5\cdot 10^{-4}B_{\text{Sch}}R^{3}=1.8\cdot 10^{9}%
\unit{G}R^{3}.$ This means that the average magnetic field at the surface of
a sphere, charged so high that its surface electric field approaches $%
1.3\cdot 10^{16}\unit{V}/\unit{cm}$ (we may think of strangelets \cite%
{strangelets} and quark stars \cite{quark stars})$,$ approaches $10^{9}\unit{%
G}.$ The magnetic moment (\ref{infiniteB}) of the ball of $R=1\unit{cm}$
and $Ze=$ $1\unit{C}$ is now $M\simeq 1.6\cdot 10^{21}\mu _{e}\simeq 15\unit{%
G}\cdot \unit{cm}^{3}.$ It is subject to a force of $2.3\cdot 10^{4}$ kG if
placed in a magnetar field $B\simeq 10B_{\text{Sch}}$.

\section*{Acknowledgements}

A. acknowledges support of FAPESP under the Contracts No. 2013/00840-9
and No. 2013/16592-4. He is thankful to Department of Physics of the University of
Florida for the kind hospitality and to Professor John Klauder for his warm
reception there. G. thanks CNPq and FAPESP for permanent support. S. acknowledges the support of FAPESP, Contract No.
2011/51867-9, and of RFBR under the Project No. 14-02-01171. He also thanks USP
for kind hospitality in Sao Paulo, Brazil, where this work was partially
fulfilled. The authors are thankful to C. Costa for discussions.

\end{document}